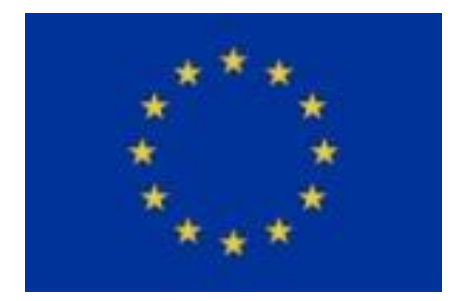

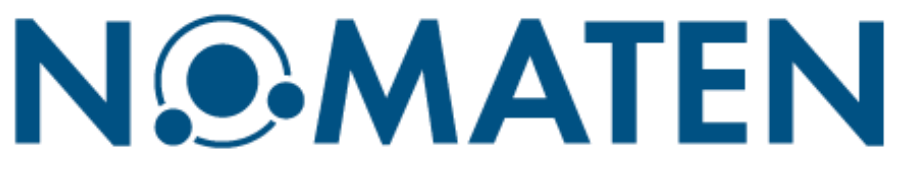

This work was carried out in whole or in part within the framework of the NOMATEN Centre of Excellence, supported from the European Union Horizon 2020 research and innovation program (Grant Agreement No. 857470) and from the European Regional Development Fund via the Foundation for Polish Science International Research Agenda PLUS program (Grant No. MAB PLUS/2018/8), and the Ministry of Science and Higher Education's initiative "Support for the Activities of Centers of Excellence Established in Poland under the Horizon 2020 Program" (agreement no. MEiN/2023/DIR/3795).

The version of record of this article, first published in Ceramics International, Volume 51, Issue 10, April 2025, Pages 12918-12931, is available online at Publisher's website: https://doi.org/10.1016/j.ceramint.2025.01.134

# High-temperature behavior of amorphous alumina coatings: Insights from *in-situ* nanoindentation and X-ray diffraction studies

A. Zaborowska[a,*)], Ł. Kurpaska[a)], M. Zieliński[a)], Q. Xu[b,*)], E. Wyszkowska[a)], J. O'Connell[c)], J.H. Neethling[c)], F. Di Fonzo[d,e)], M. Frelek-Kozak[a)], S. Papanikolaou[a)], R. Diduszko[a)], J. Jagielski[f)]

*[a)] National Center for Nuclear Research, NOMATEN CoE MAB+ Division, A. Soltana 7, 05-400 Otwock-Swierk, Poland*
*[b)] Institut Pprime, CNRS - Universite de Poitiers - ENSMA, UPR 3346, Departement de Physique et Mecanique des Materiaux, Bvd M. et P. Curie, SP2MI, BP 30179, 86962 Futuroscope Chasseneuil Cedex, France*
*[c)] Centre for High Resolution Transmission Electron Microscopy, Physics Department, Nelson Mandela University, University Way, Summerstrand , 6031, Port Elizabeth, South Africa*
*[d)] Center for Nano Science and Technology @PoliMi, Istituto Italiano di Tecnologia, Via Giovanni Pascoli 70/3, 20133 Milano, Italy*
*[e)] X-nano S.r.l, Via Rubattino 81, 20134 Milano, Italy*
*[f)] National Center for Nuclear Research, A. Soltana 7, 05-400 Otwock-Swierk, Poland*

* *corresponding author: agata.zaborowska@ncbj.gov.pl, qinqin.xu@univ-poitiers.fr*

## Abstract

Further development of nuclear power plant technology relies heavily on materials' durability under operating conditions. Estimating the materials' performance in the *operando* tests is crucial. In this paper, the mechanical behavior of thin amorphous nuclear-dedicated $Al_2O_3$ coatings deposited by pulsed laser deposition was investigated by nanoindentation over the temperature range of 25-650 °C. Experimental nanomechanical analysis was supported by MD simulations. The results indicate that the hardness of the amorphous coating experiences a gradual, constant decrease with temperature, while the Young modulus value remains constant in the whole temperature range. Observed phenomena confirm the increasing plasticity of the material and it is postulated to be related to the bond-switching mechanism that accelerates at high temperatures. The post-mortem transmission electron microscopy characterization confirmed that the loaded material was non-crystalline over the entire range of the indentation temperatures. The thermal stability of the structure was further studied *in-situ* up to 1050 °C by X-ray diffraction. The implemented methodology allowed us to follow the dynamic process of phase transitions occurring in the material above 650 °C. First, thermally activated crystallization was observed at 700 °C. Intermediate alumina phases were present up to 950 °C, while above this temperature, exclusively the thermodynamically stable $\alpha$-$Al_2O_3$ was observed. The *in-situ* high-temperature characterization of the evolution of thin films boosts the understanding of the application limits of the coating systems at elevated temperatures. The added value is that the paper demonstrates the potential usefulness of combining high-temperature techniques to characterize the complete behavior of thin films at elevated temperatures.

Abbreviations:

HT – high temperature

RT – room temperature

## 1. Introduction

Developing materials for extreme applications is very challenging as it requires consideration of many factors and complex relations that can affect the system's performance. This is best exemplified by nuclear materials that must retain their full functionality in the most demanding conditions, including high temperature (HT), high levels of radiation, mechanical stresses, and corrosive environment [1–3]. Materials for such applications are required to display appropriate characteristics, particularly in terms of resistance to environment-induced property degradation.

To ensure that the given material will retain its durability within specified tolerances throughout its service lifetime, tests simulating the influence of operating conditions on material properties are performed. Concerning specifically high-temperature applications, assessment and verification of permanence of material performance at temperatures that match its actual operating conditions are crucial. Thus, tests conducted at the temperature of interest are much more relevant and valuable than room-temperature experiments [4]. Various high-temperature methods have developed significantly over the last few decades, thus becoming a powerful tool for investigating specific materials properties under non-ambient conditions. Depending on the specific designation, high-temperature examination may involve e.g. thermal, electrical, optical, corrosion, and mechanical tests [5–8].

As for the latter, traditionally, mechanical tests are performed according to well-established procedures set out by international standards. This applies to both room and elevated temperature measurements. The issue arises when one doesn't have the adequate quantity of the material to meet sample dimension requirements imposed by the standards. This is very often the case in post-irradiation examinations or in preliminary studies in the field of nuclear materials science. For example, newly developed material grades are often manufactured on a laboratory scale, which means that the material available for testing is very limited. For the research and development of new advanced nuclear materials systems, accessibility of micro-scale mechanical characterization techniques is essential [9]. This is one of the main driving forces for developing miniature test specimens. However, the task becomes even more challenging when one needs to characterize a volume of the order of $\mu m^3$. The key method that makes it possible is nanoindentation. Nanoindentation stands out among other available small-scale techniques for mechanical testing (e.g., micro-compression, micro-bending, or micro-tensile testing) due to the simplicity of the sample requirements, cost-effectiveness, and relative ease of use [10,11]. But also – or perhaps even most importantly – nanoindentation allows to obtain mechanical property information from thin films and coatings, which might not be feasible using other techniques [4,11–15].

This is the case with PLD-grown thin amorphous alumina coatings dedicated to new advanced nuclear energy systems such as Lead-Cooled Fast Reactors (LFR). PLD-grown alumina is a main candidate material for the fuel cladding that will be operated at 550 ºC (600 °C in hot spots) [16]. The power of alumina is that, in contrast to available metallic materials such as chromia forming austenitic stainless steels (e.g. 316L, 15-15 Ti) or ferritic-martensitic (F-M) steels (e.g. T91), it is corrosion resistant in the liquid lead and lead-bismuth eutectic (LBE) at a temperature above 500 °C [17–22]. A significant amount of data on the room temperature nanoindentation behavior of the material was published in the literature in the last decade. For example, the technique was utilized to characterize 1 μm thick coatings in the as-deposited state and to evaluate the influence of radiation on the material performance [22–26]. Knowledge gained so far covers the issue of the sole irradiation effect (RT experiment [23,26]) and the combined effect of high temperature and irradiation (600 °C experiment [22,24,25]) on the PLD-grown coatings mechanical performance. Published data revealed that room temperature irradiation (up to 50 dpa) has virtually no impact on the nanomechanical properties of the material [23,26]. At the same time, for the same dpa level, high-temperature irradiation significantly increases the coating hardness [25]. Accompanied structural analysis showed that observed hardness increase follows from radiation-induced crystallization of the initially amorphous alumina phase [25]. The phenomenon occurring in the material is considered to be the result of the synergetic effect of high temperature and radiation damage. Hence, the lower the irradiation temperature, the higher the dose threshold for the onset of crystallization. All these findings are extremely important, but there is a piece missing. To the best of the authors' knowledge, despite the growing interest in PLD-grown alumina, no information about the sole effect of high temperature on the performance of amorphous PLD-grown alumina can be found in the literature.

Meanwhile, nanoindentation measurement capabilities developed significantly, enabling measurements to be carried out under non-ambient conditions [4,10,13,27–29]. D. Beake, in his reviews [4,13], presented state-of-the-art *in-situ* high-temperature nanomechanical testing of materials dedicated to extreme environments. The review shows that although certain challenges related to the technique still exist (e.g. thermal control management, limited durability of indenter material, no established testing protocol) many published studies successfully applied the methodology to measure the mechanical response of various materials at high temperatures [28,30–33], including thin coatings [34,35]. In our research, the technique was utilized to broaden the state-of-the-art data on PLD-grown alumina behavior with *in-situ* high-temperature information collected up to 650 °C. We confirmed that theoretical predictions of amorphous alumina nanoscale plasticity [36] extends to the microscale at room temperature [26] are correct, and brittleness of this material could be limited further at high-temperature. This confirms that amorphous alumina oxide show significant potential to be used as a light, high-strength, and damage-tolerant engineering material which could be applied in Gen. IV applications.

To gain a complete picture of material behavior at high temperatures, the mechanical investigation was supplemented with an *in-situ* structural X-ray diffraction investigation (up to 1050 °C). It is known that the transformation of the amorphous $Al_2O_3$ into stable $\alpha$-$Al_2O_3$ can be triggered solely by high temperature [37]. In nature, aluminum oxide (alumina, $Al_2O_3$) exists in several polymorphs, including stable $\alpha$ (corundum) and various metastable forms (e.g. $\gamma$, $\sigma$, $\eta$, $\chi$, $\kappa$, $\theta$) [19]. Depending on the material processing route, the formation of different metastable structures and the various sequences of phase transformations towards the stable $\alpha$-$Al_2O_3$ may be observed [38–40]. For this reason, the presented study aimed to provide specifics on the thermal evolution of PLD-deposited $Al_2O_3$. Nanoindentation results were analyzed in the context of structural output, thus further contributing to the fundamental understanding of the basis of materials' mechanical behavior.

To summarize, the only way to understand the whole picture of the in-service behavior of alumina is to investigate its properties in different conditions and study them independently. Meanwhile, no information about the sole effect of high temperature on the performance of amorphous PLD-grown alumina can be found in the literature. Such information would be helpful not only to understand better the material's behavior subjected to nuclear operating conditions but also to find out how the material would perform in potential non-radiation high-temperature applications. In this study, thin PLD-grown alumina coatings dedicated to nuclear applications were analyzed for their high-temperature properties. To the best of the authors' knowledge, the presented data represent the first reported strictly high-temperature information on amorphous alumina.

## 2. Material and methods

### 2.1. Materials preparation

5 μm thick alumina coatings were deposited at RT on 0.9 mm thick AISI 316L stainless steel by PLD (Pulsed Laser Deposition). It was reported that PLD-grown coatings generally reproduce the roughness of the substrate [41]. Therefore, prior to the manufacturing process, steel samples were ground with SiC abrasive papers (up to 1200 grade) and finely polished with colloidal silica suspension (0.06 μm grain size). More details on the deposition process can be found in the reference [41].

### 2.2. *In-situ* high-temperature nanoindentation

In this study, the nanomechanical properties of alumina were investigated *in-situ* at elevated temperatures. The measuring range was from RT to 650 °C, in incremental steps of 100 °C. The upper-temperature range was limited by the equipment capabilities but sufficient to go beyond the foreseen operating conditions (600 ºC in hot spots). The instrument utilized was the NanoTest Vantage System from Micro Materials. Our experimental methodology follows the guidelines indicated in [35]. The measurements were performed in depth-control mode. The target depth was set to 300 nm, corresponding to forces ranging from 15 to 21 mN, depending on the temperature. According to the 10% rule [42], 300 nm indentation depth results in the plastic zone about 3 μm deep into the material. Given the 5 μm thickness of the investigated coating, the plastic stress field should remain in the coating, which means that the mechanical property data were extracted exclusively from the alumina (no substrate contribution). Hardness and Young's modulus values were determined using the method developed by Oliver and Pharr [43]. For each temperature test point, ten indentations were performed.

Since nanomechanical measurements at higher temperatures are extremely sensitive and experimental protocols are not yet well established, the test design was the subject of particular attention. The following is the list of key issues that were taken into consideration.

(a) Environment control

The instrument features a controlled atmosphere chamber, allowing operation in a protective gas atmosphere. In this work, the chamber was purged with high-purity argon gas, thus restricting the oxygen level in the test environment. An overpressure was established, and the oxygen content level was below the detection limit of the sensor, which was expected not to affect the corrosion behavior of alumina.

(b) Indenter choice

The choice of indenter material for high-temperature measurements is of vital importance. To limit tip degradation, indenter material within the temperature of interest must satisfy several criteria: hardness significantly higher than of tested material and low chemical reactivity with the sample [4,10,44]. The review studies on high-temperature nanoindentation state that diamond and cubic boron nitride (c-BN) are the most commonly used materials for high-temperature indenters. Due to the highest hardness, when possible, the use of diamond is recommended [44]. However, it is well known that diamond oxidizes in air at temperatures above 400-500 °C [4,10,35,44]. As mentioned above, in the system utilized, the sufficiently low oxygen content level could not be strictly controlled. Therefore, in this study, c-BN was chosen as a material for the Berkovich indenter. The advantage of this material is its chemical inertness. However, c-BN is not as hard and durable as diamond, which may lead to rapid tip wear at high temperatures. Consequently, the Diamond Area Function (DAF) was recalculated after the 250 °C step due to the significant number of tests conducted at RT, 150 and 250 °C, as well as the system's cooling between the 250 °C measurements and subsequent higher temperature tests. This recalibration ensured that any wear sustained during the initial tests (RT, 150 °C, and 250 °C) was accounted for, enabling the use of an updated DAF for measurements at higher temperatures. One should mention that the used indenter tip was brand new. Therefore, we assume that errors owing to the tip bluntness were minimized.

(c) Thermal drift

It has been repeatedly shown that it is crucial to minimize thermal drift to achieve high-quality, accurate high-temperature nanoindentation data. Thermal drift is a time-dependent error in displacement measurement resulting from thermal expansion occurring in the system [10]. Thermal drift consists of two components: frame drift and contact drift [10]. The former is the effect of temperature variations in the system’s frame and can be simply mitigated by including a stabilization period [4,10]. In this study, to stabilize the frame, prior to measurements, the sample was held at the desired testing temperature for a minimum of 4 hours.

The latter drift occurs due to the heat flow between the sample and the indenter at the contact and can be minimized by ensuring that the temperature of these components is precisely matched [4,10,12,35,44]. System configuration plays an important role in achieving an isothermal contact. In the NanoTest Vantage System from MML, the sample and the indenter are independently actively heated, which allows to reduce thermal drift during the measurements. Furthermore, in addition to the integrated control thermocouples for the indenter and sample stage, the system is equipped with an auxiliary thermocouple, which, placed directly on the sample surface, indicates true contact temperature. Another thermal drift mitigation strategy consisted in using a shroud around the sample holder. This simple procedure allows for stabilizing the temperature in a certain chamber volume, which is very important when the testing temperature exceeds 350-400 °C.

As discussed above, the temperature's true value on the sample's surface was monitored by the auxiliary thermocouple. This proved to be essential, as due to the low thermal conductivity of the coating, a significant thermal gradient in the volume of the sample was observed. However, it is not feasible to implement an analogous solution for the indenter; the thermocouple that controls its temperature is always shifted in relation to the contact area. Due to the existing thermal gradient, the exact value of the indenter tip's temperature is, therefore, unknown. To adjust the temperature of the indenter tip to the temperature of the sample, at each temperature, a series of test indentations were performed, and based on the recorded thermal drifts, the indenter set temperature was tuned until the thermal drift was minimized to the acceptable level. A schematic representation of the high-temperature nanoindentation setup is shown in Fig. 1.

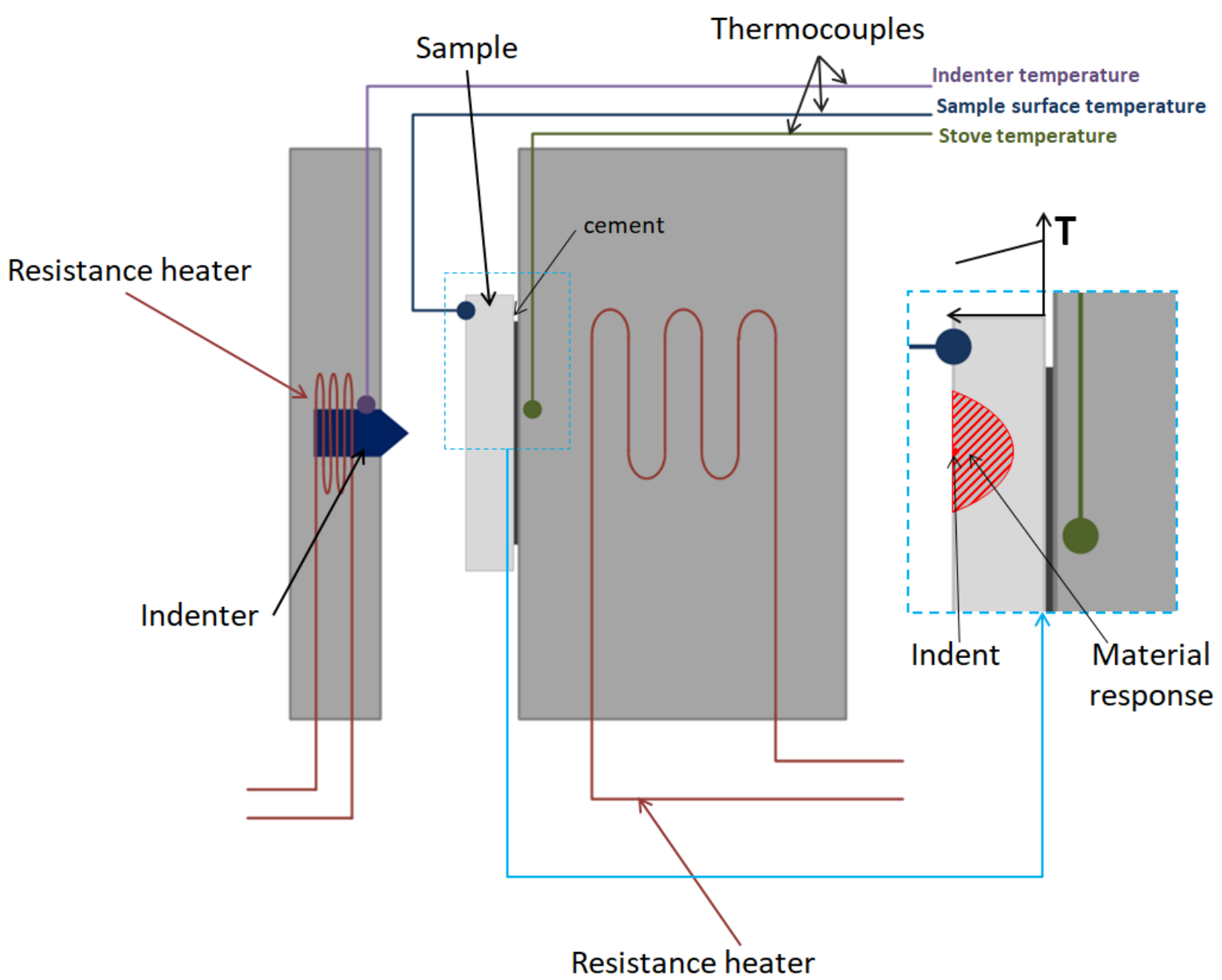

Fig. 1. Schematic representation of high-temperature nanoindentation setup.

Our nanoindentation methodology owes a lot to the one proposed by Rebelo De Figueiredo et al. [35]. To further minimize the temperature heat transfer at the contact, before each single measurement, a second thermal stabilization step was included – 180 s of thermalization period at the distance between the sample surface and indenter tip not exceeding 2 µm. Each single indentation cycle included: loading with the rate of 2 mN/s, 30 s of dwell period at maximum load, and unloading with the rate of 5 mN/s. The thermal drift correction data were collected from the hold period of 60 s at 10% of the maximum load applied while the thermal drift was calculated over the final 60% of the hold period. The goal of this was to account for the creep recovery, which is significant at at highest temperature. Described methodology aligns well with ISO 14577 standard.

The highest recorded thermal drift in this study was 0.65 nm/s at 550 °C. For most indents below 350 °C, it didn't exceed 0.35 nm/s, and for all other temperatures mostly didn't exceed 0.50 nm/s. It should be noted that some measurements with a thermal drift close to zero were registered.

2.3.Post-indentation structural characterization

To link the mechanical information with potential changes in the coating structure, the indented sample was examined by Transmission Electron Microscopy (TEM) and Grazing Incidence X-Ray diffraction (GIXRD) after cooling down to RT.

### 2.3.1. Transmission electron microscopy

Samples for TEM observations were fabricated by sectioning a thin electron transparent lamella. Lamella was cut by conventional Focused Ion Beam (FIB) technique using a FEI Helios Nanolab 650. Final thinning of the lamella was performed with 5 keV Ga ions followed by a 2 keV Ga polish. Electron transparent lamellae were analyzed in a JEOL ARM 200F operating at 200 kV and equipped with an Oxford Xmax100 EDS detector. Probe current density was sufficiently low to avoid any obvious beam induced structural modification. The selected area electron diffraction (SAED) analysis was performed in TEM mode using parallel illumination.

### 2.3.2. Grazing incidence X-Ray diffraction

Diffraction patterns were acquired in the 2θ range from 10° to 145° using the parallel X-Ray beam obtained with a Göbel mirror at the Bruker D8 Advance diffractometer. The X-Ray tube had a Cu anode, 40 kV accelerator and 40 mA current ($\lambda_{CuK\alpha1}$ = 1.5406 Å). The grazing incidence beam geometry was chosen to maximize the contribution of the $Al_2O_3$ signal in the detected diffraction pattern, thus limiting the acquired scattering of the substrate. The incidence angle ω was set to 0.7° or 1.15°. A 0.1 or 0.2 mm fixed primary slit, the 5 mm vertical primary beam mask, the 0.2° secondary equatorial Soller (a.k.a. Parrish-Hart analyzer slits), and 2.5° primary and secondary axial Soller slits were used. A LYNXEYE XE-T detector working in the high energy resolution ($\Delta E < 380$ eV at 8 keV, energy window optimized for $CuK_\alpha$ lines) and 0D modes without Ni-filter was used. The Bruker DIFFRAC.EVA program with the database of reference crystal structures ICDD PDF4+ 2022 [45] were used for qualitative phase analysis.

## 2.4. *In-situ* high-temperature X-Ray diffraction

Diffraction patterns acquisition during the *in-situ* high-temperature measurements was also performed using the parallel X-Ray beam obtained with a Göbel mirror and the grazing incidence beam geometry (ω = 1.25°). The diffractometer configuration was the same as described in section 2.3.2, but only the 0.2 mm primary slit and no beam mask were used.

High-temperature measurements were carried out in a HTK 1200 N chamber by Anton Paar, with resistive heating (Kanthal wire) and temperature control based on read-outs from a thermocouple sensor. The Kapton$^{TM}$-graphite window is divided in 2 parts and the pillar in between them limited the covered 2θ range to 83°. The region above 96° included a large number of weak overlapping peaks not useful for the analysis. Temperature ranged from ambient to 1050 °C with the following values being chosen: 25 °C and 10 steps with the increment of 50 °C between 600 °C and 1050 °C. The dwell time at each temperature was not shorter than 18 hours. An ambient air atmosphere was used. The position of the sample surface (its height and tilt) was realigned at each temperature with respect to the goniometer axis and beam direction at $\theta = 0°$.

### 2.5.High-temperature nanoindentation simulations - Molecular Dynamics (MD) simulations

Molecular Dynamics (MD) simulations were performed to assist the investigation of the material properties of amorphous $Al_2O_3$ samples under nanoindentation at elevated temperatures, using LAMMPS [46]. All atomistic simulations were carried out using the 2/3-body Vashishta potential used to describe the cohesive energy, elastic constants, bulk modulus, and melting temperature of the corundum phase. A cutoff of 6.0 Å was used as in the original study of Vashishta and collaborators [47]. Fig. 2 shows the amorphous alumina sample under nanoindentation. It was created with a cell size of 30.2nm × 30.3 nm × 35.9 nm (length × height × width). Amorphous alumina was fabricated by cooling the melt. The melting process started with a crystalline $\alpha$-$Al_2O_3$ slab sample of approximately 3,300,000 atoms with an NPT ensemble under ambient pressure at 4000 K for 50 ps. The system was cooled to different temperatures from the liquid state with a rate of 0.15 K/ps. Finally, the glass was relaxed at a targeted temperature under atmospheric pressure for 10 ps. Periodic boundary conditions were applied in all directions. Periodic boundary conditions were applied in all directions. Several simulated melt-and-quench studies that are comparable with this work exist. For example, Momida et al. [48] simulated the initial configuration with alpha phase corundum structure, which is similar to our MD simulations, while Giacomazzi et al. [49] used the kappa phase as the initial one. After the fabrication stage, the samples were divided into three sections in the z direction to set up boundary conditions along its depth, dz: 1) a frozen section with a width of approximately 0.02×dz, which was used for stability of the numerical cell; 2) a thermostatic section at approximately 0.08×dz above the frozen section, which was used to dissipate the heat generated during nanoindentation; and 3) a dynamical atoms section, where the interaction with the indenter tip modifies the surface structure of the samples. In addition, a 5 nm vacuum section was included at the top of the sample as an open boundary.

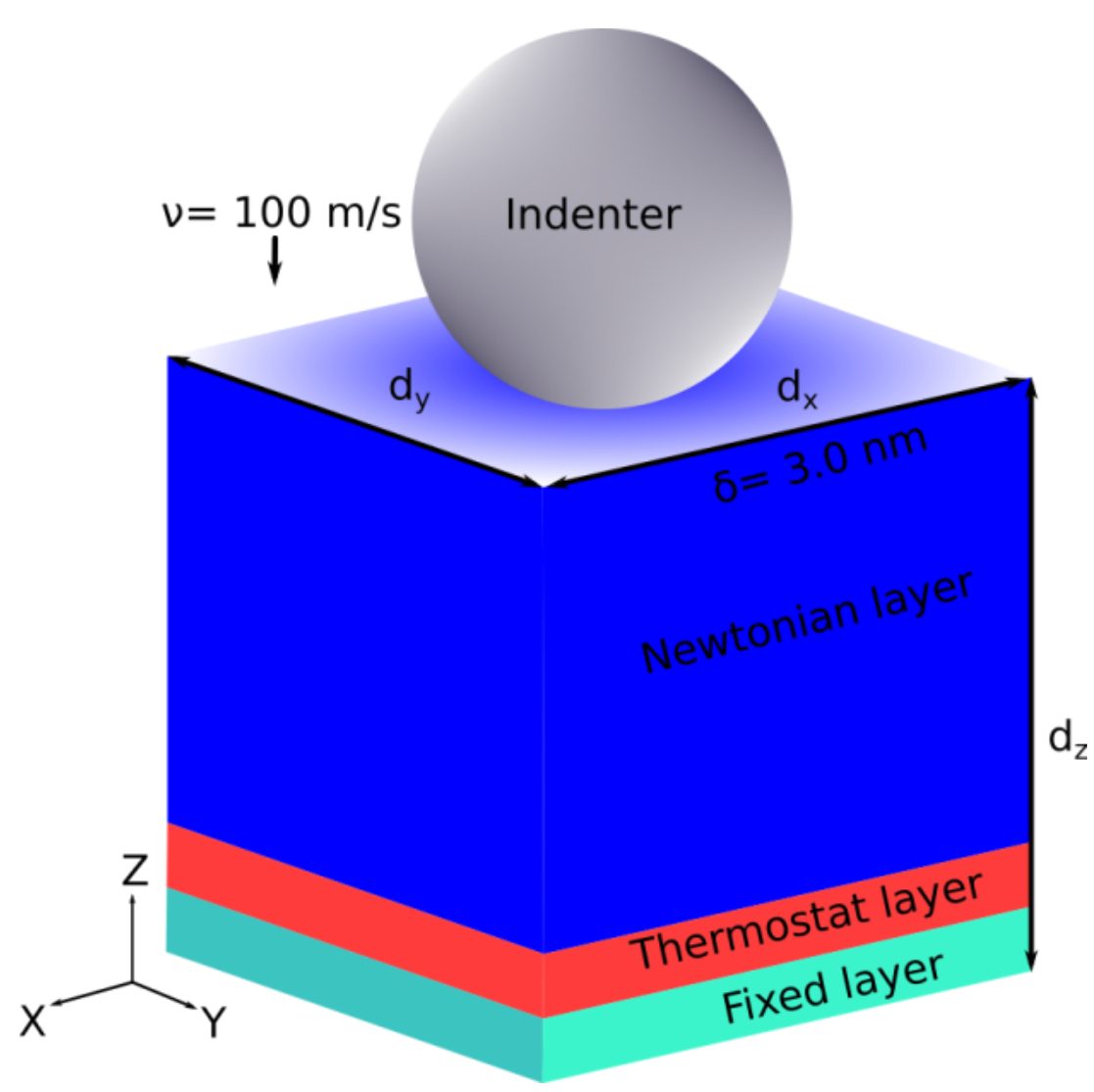

Fig. 2. Schematic of the MD simulation model for the nanocontact between a spherical indenter tip and alumina sample (cell size 30.2nm × 30.3 nm × 35.9 nm).

The indenter tip was considered a non–atomic repulsive imaginary (RI) rigid sphere with a force potential defined as $F(t) = K(\vec{r}(t) - R)^2$, where K = 936 eV/Å$^3$ is the force constant, and $\vec{r}(t)$ is the position of the center of the tip as a function of time, with radius R = 7 nm. We applied MD simulations using an *NVE* statistical thermodynamic ensemble and the velocity Verlet algorithm to emulate an experimental nanoindentation test. Periodic boundary conditions were set on the x and y axes to simulate an infinite surface. At the same time, the z orientation contained a fixed bottom boundary and a free top boundary in all MD simulations. Here, $\vec{r}(t) = x_0\hat{x} + y_0\hat{y} + (z_0 \pm vt)\hat{z}$, with $x_0$ and $y_0$ as the center of the surface sample on the xy plane. The initial gap between the surface and the indenter tip, $z_0$ = 0.5 nm, moves with a speed of v = 100 m/s. The center of the indenter tip was randomly changed to 10 positions to consider statistics in our results, resulting in 150 MD simulations. Each process was performed for 125 ps with a time step of Δt = 1 fs. The maximum indentation depth was chosen to be 3.0 nm to avoid the influence of boundary layers in the dynamical atom region.

The hardness of the indented sample was calculated as in [50] by computing the P − δ curve with the Oliver and Pharr method [43], following the fitting curve to the unloading process curve as:

$$P = P_0(\delta - \delta_f)^m \qquad (1)$$

where P is the indentation load, $\delta_f$ is the residual depth after the whole indentation process, and $P_0$ and m are fitting parameters. Thus, the contact pressure is computed as:

$P(\delta)/A_c$ (2)

where P is the load as a function of the indenter displacement and $A_c(\delta)=\pi(2R-\delta)\delta$ is the contact area at load P, considering the indenter displacement, δ, and tip radius, R [51]. The effective elastic modulus is calculated by using Oliver-Pharr method [43], where the projected contact are is obtained by considering the maximum depth as $\delta_c = \delta_{max} - \varepsilon * P_{max}/S$ where $\varepsilon = 0.75$, which refers to the parameter of spherical indenter shape. The unloading stiffness S is calculated as:

$$S = (dP/d\delta_{\ \delta=\delta max}) = mP_0 (\delta_{max} - \delta_f)^{m-1} \quad (3)$$

Then, the Young modulus is $E_Y$ is computed as:

$$(1-\nu^2)/E_Y = 1/E_r - (1-\nu_i^2)/E_i \quad (4)$$

where ν and $\nu_i$ are the Poisson's ratio of sample and the indenter, respectively. $E_i$ is the Young modulus of the indenter tip and is considered to be infinitively large. Thus the effective elastic modulus is expressed as $E_r = \sqrt{\pi/A_c(\delta_c)}S/2\beta$ with β=1 for spherical indenter tip. We reported the following values that were computed at the maximum indentation depth: maximum force, $P_{max}$ in μN, the projected contact area, $A_C$ in $nm^2$, the stiffness, S in units of N/m, and Hardness in units of GPa. The Hertz model $F(\delta) = 4/3\ E_r r^{1/2} \delta^{3/2}$ where $E_r$ is a reduced Young's modulus and δ the indenter depth is used to identify the pop in event where the transition from elastic to plastic deformation occurs.

## 3. Results

### 3.1. High-temperature nanoindentation - experiment

High-temperature nanoindentation measurements were performed to study the evolution of the mechanical properties of the investigated material over LFR foreseen operating temperatures. The load-displacement behavior of a-$Al_2O_3$ coating within the testing temperature range is presented in Fig. 3. It can be observed from the maximum load segment of the curves that creep deformation becomes increasingly prominent with temperature. This result suggests that investigated material increases its ductility with increasing temperature. The shape of the unloading curves, free of artifacts, suggests that the dwell time was adjusted correctly (the deformation generated under the indenter tip develops uniformly, no burst-like-events were recorded) and accurate fitting for reduced elastic modulus calculations can be performed.

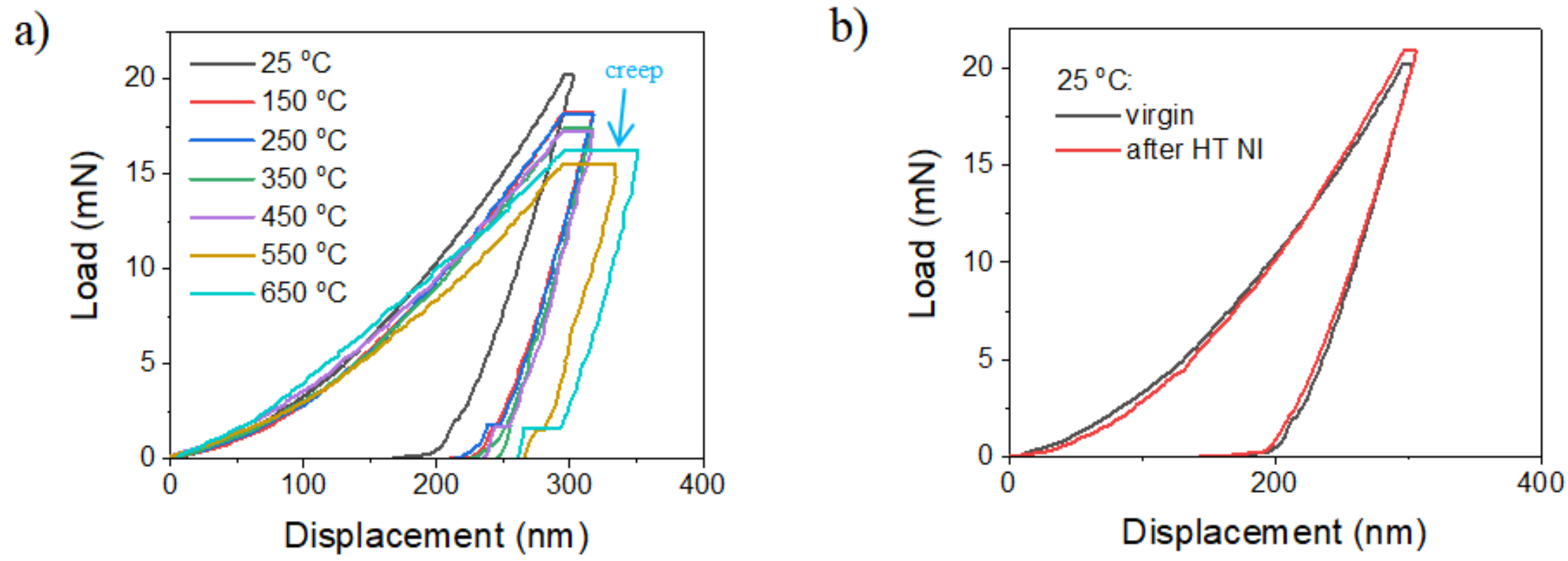

Fig. 3. Representative load-displacement curves for a-$Al_2O_3$ coating measurements (a) in the range of 25 to 650 °C and (b) at 25 °C before and after HT nanoindentation.

The corresponding mechanical property values are plotted as a function of temperature in Fig. 4. The initial nanohardness of the as-deposited coating is 9.6 ± 0.4 GPa, which is in complete agreement with the previously reported values [23,41]. As shown in Fig. 4, the hardness of the alumina decreases with increasing temperature. At 550 °C, the average hardness drops to 5.2 ± 0.2 GPa, representing 54 % of the room temperature value. According to the structural analysis reported in the following sections, the material subjected to load was non-crystalline over the entire range of the indentation temperatures. Thus, we look into the high-temperature mechanical behavior of the amorphous material. Reported softening phenomenon fits well with previous literature findings for crystalline alumina [35,52–55]. In Fig. 4, the RT mechanical values for the sample after the high-temperature nanoindentation were plotted with red symbols. Compared to the initial value, the hardness of the thermally annealed sample was slightly higher (11.9 ± 0.8 GPa). A reasonable explanation was found and discussed along with the TEM results (see the next section). Interestingly, no significant effect of the temperature on the materials' Young's modulus was found. Previously reported research on Young's modulus temperature dependence of crystalline alumina [35,52–55] indicates that the material's stiffness decreases with the temperature. Thus, our study suggests that the properties of the studied amorphous material differ from those of crystalline bulk samples reported earlier in the literature.

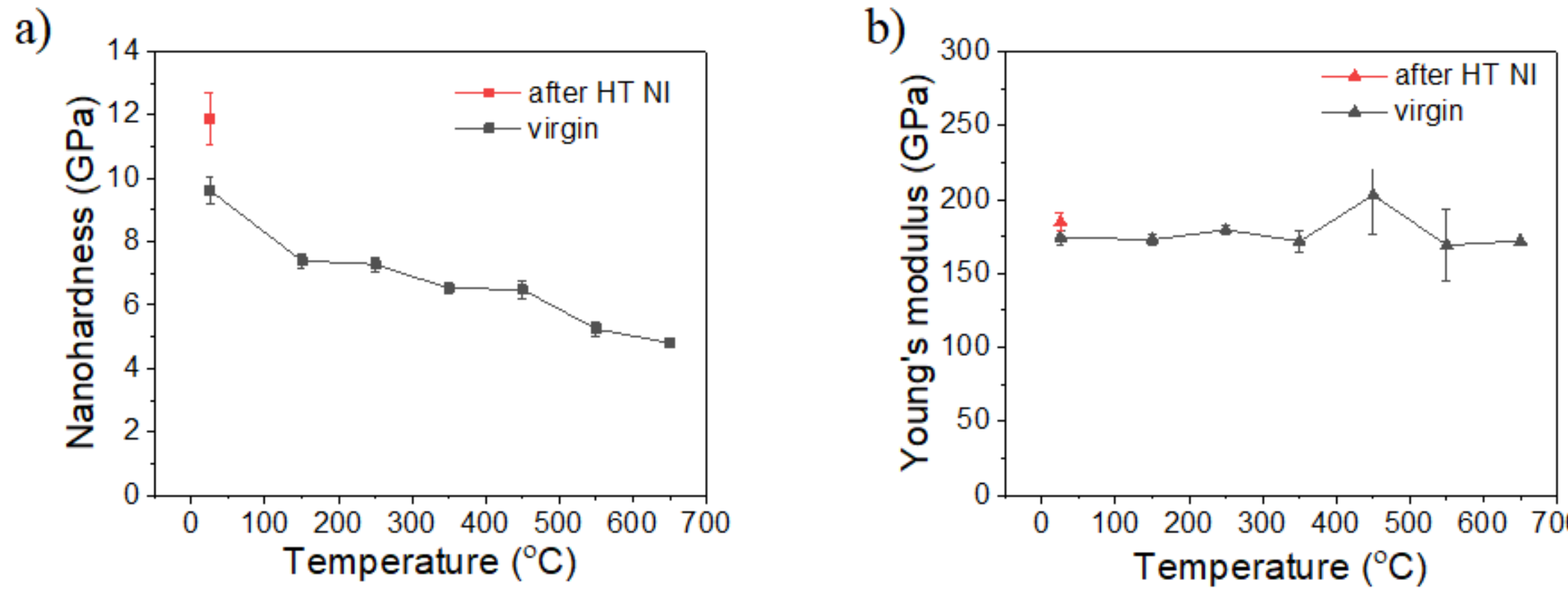

Fig. 4. Temperature dependence of the (a) nanohardness and (b) Young's modulus of a-$Al_2O_3$ coating.

In addition, data on elastic-plastic work was extracted from the experimental measurements. Notably, the plasticity of the material was observed to increase with rising temperature, as depicted in Fig. 5. This is best exemplified by the microhardness dissipation parameter (MDP), which is defined as the plastic work divided by the total deformation work measured during indentation [41,56,57]. MDP can be understood as a parameter that can be used to characterize the ability of the coating to dissipate energy. The results may suggest that the durability of the coating improves with the temperature.

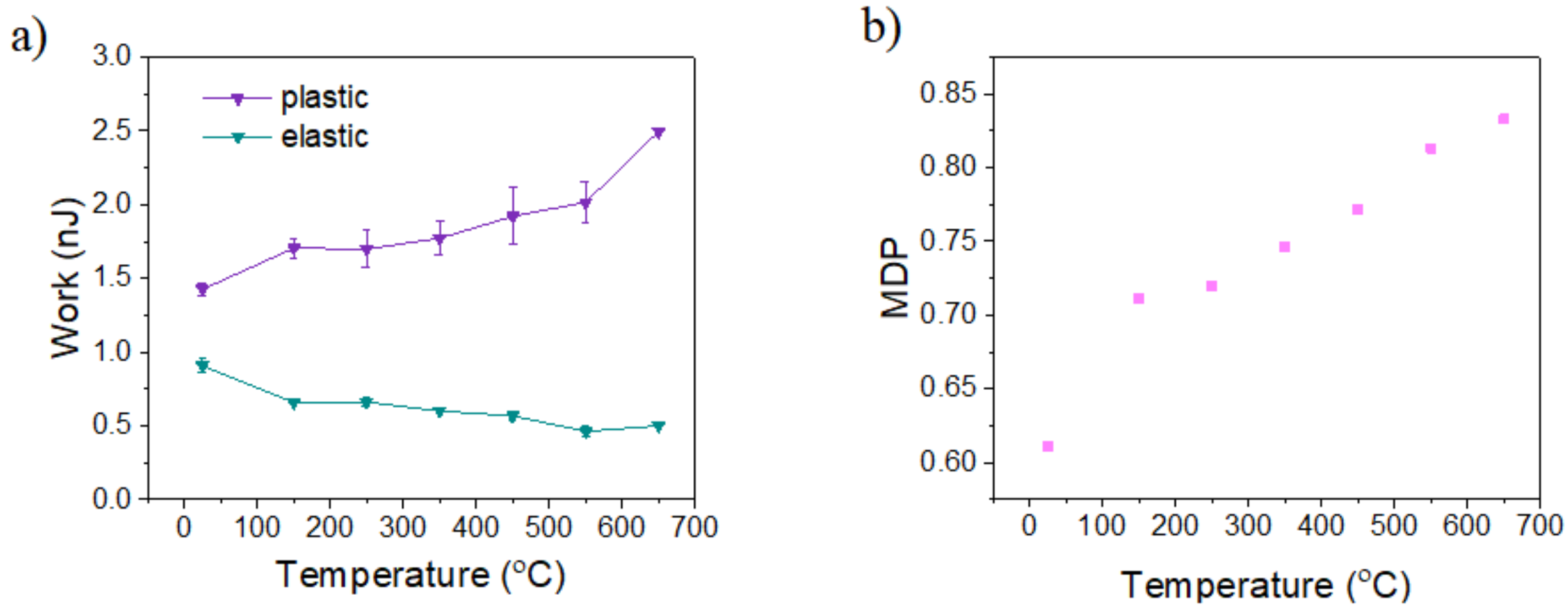

Fig. 5. Temperature dependence of the (a) work and (b) MDP of a-$Al_2O_3$ coating.

## 3.2. Post-indentation structural characterization

### 3.2.1. Transmission electron microscopy (TEM)

The TEM technique was used to examine the microstructure of the sample after high-temperature nanoindentation. Images of a cross-section through the coating are presented in Fig. 6. The entire layer is visible in Fig. 6b with the steel substrate at the top, which is around 4.9 μm thick.

A BF TEM image of the steel-$Al_2O_3$ interface is presented in Fig. 6c. One can observe that crystallization in the interfacial region occurred. The crystallized alumina layer comprises columnar grains grown perpendicular to the substrate surface. The thickness of this region varied from 300 to max. 750 nm. The documented crystallization aligns with the findings of the high-temperature XRD measurements, described hereinafter, revealing that the crystallization temperature for the amorphous phase lies between 650 and 700 °C. As explained in the *Material and methods* Section, the nanoindentation system heats the sample from the bottom, causing a thermal gradient across the material cross-section. For the 650 °C indentation (temperature at the sample surface), the heater temperature was set to 720 °C. Therefore, the bottom of the sample was at higher temperature in relative to the top, and the temperature in this region exceeded the crystallization point determined by XRD (details presented in the following section). This explains the localization and columnar nature of the crystalline alumina region located near the metallic substrate.

It should be noted that according to the 10% rule, which states that the extent of the plastic zone is 10 times the indentation depth [42] – this layer has no contribution to the mechanical response registered. This is graphically demonstrated in Fig. 6b. Furthermore, it can be confidently postulated that the initiation of the crystallization process occurred at the final temperature measurement step (650 °C), thereby having no influence on all prior indentations.

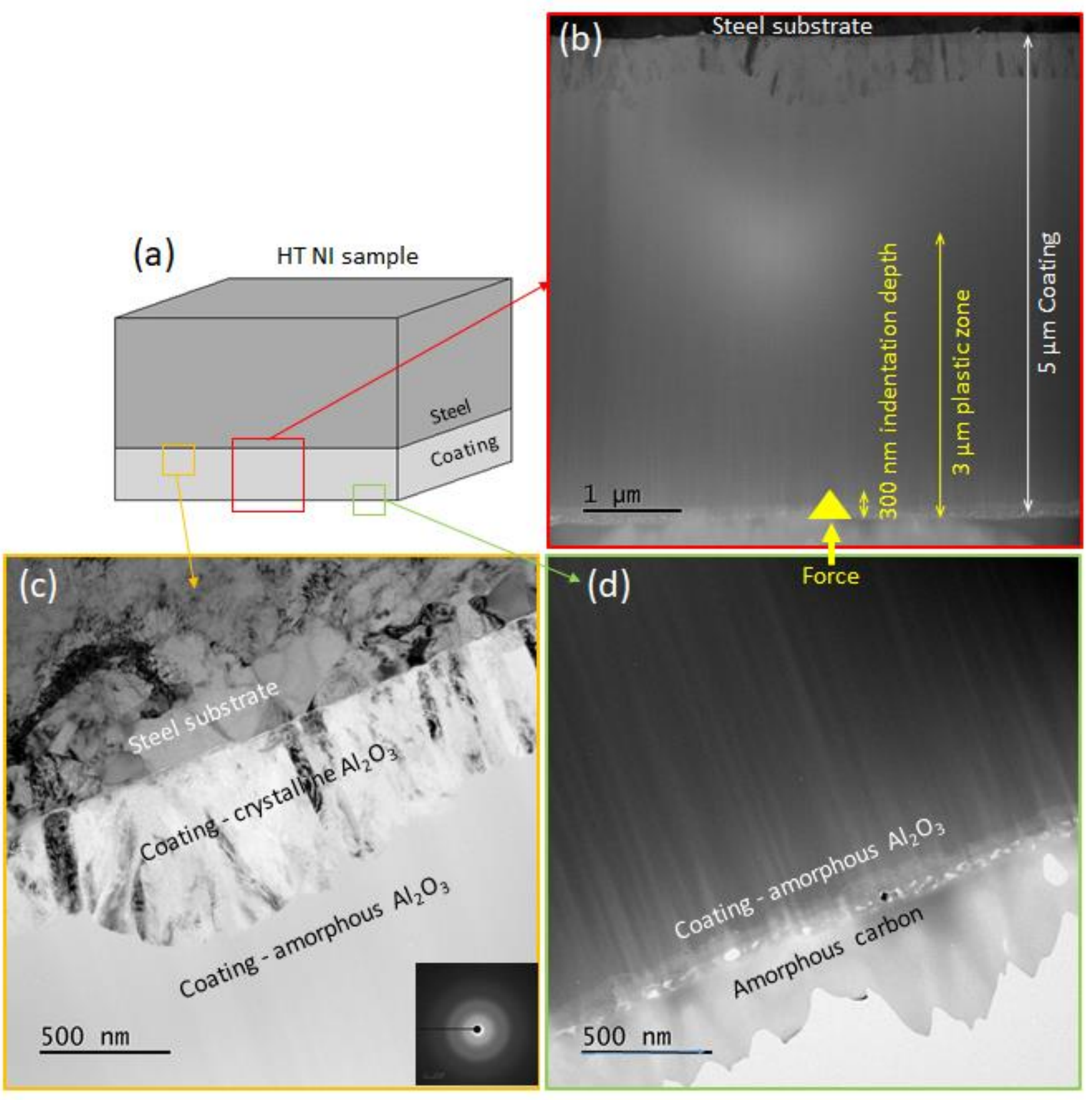

Fig. 6. Schematic representation of the (upside down) sample after HT nanoindentation (a) together with BF TEM images of cross-sections through the coating showing the entire thickness (b), steel-coating interface (c), and the top surface of the coating (d). Schematic yellow details marked on the upper-right image are for scale presentation purposes only and do not indicate the exact spot where the indent was made.

Fig. 6d gives a view of the specimen surface. The lighter "curtain" stripes with a few nm in thickness running normal to the surface are artifacts form lamella thinning and may be neglected. These are the effects of local thickness variations due to modulation of the Ga beam focus during thinning. The surface region consists of two continuous crystalline layers with a total thickness of about 100 nm. The top sublayer is very porous and consists of small crystallites identified as $Al_2O_3$. The transition sublayer below, in turn, contains a reasonable amount of phosphorus (EDX measured). As crystalline alumina yields higher hardness (~25 GPa) than amorphous (~10–12 GPa) [58], we registered a slight increase in the sample hardness after being cooled down to the RT. This near-surface layer contribution could explain this effect. More details on this will be given in the next Section.

### 3.2.2. Grazing Incidence X-Ray diffraction (GIXRD)

The results of the GIXRD measurements on the sample after HT nanoindentation are shown in Fig. 7. By varying the incidence angle, the contribution to the diffraction pattern from the very surficial layer changes according to the effective penetration depth. For ω set at 0.7° or 1.15°, 95% of information originates from a depth of approx. 3.1 μm or 5.1 μm, respectively (assuming the $Al_2O_3$ density is 3.7 g · $cm^{-3}$).

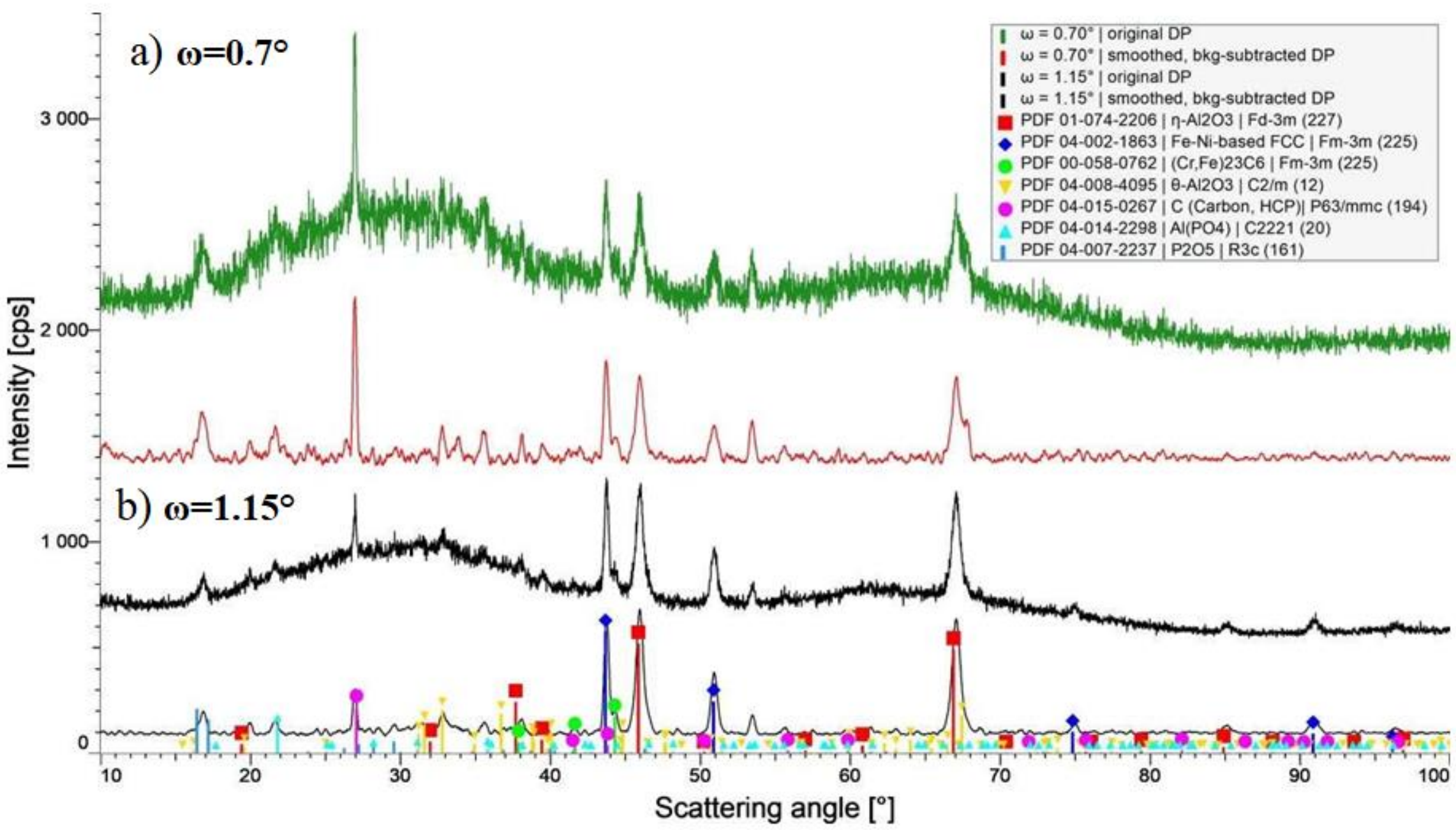

Fig. 7. Comparison of GIXD accumulated diffraction patterns raw and after smoothing and background subtraction for the incidence angle (a) ω = 0.7° (green and red curves, respectively) and (b) 1.15° (black curves). Vertical lines and symbols mark diffraction lines for the identified crystalline phases. In addition to η-$Al_2O_3$ and steel other probable phases include: i) present in the steel substrate $(Cr, Fe)_{23}C_6$; ii) present at the substrate-coating interface $Fe_2NiO_4$; iii) minor phase present in the $Al_2O_3$ coating – monoclinic θ-$Al_2O_3$; iv) present at the surface – elemental carbon (C), $AlPO_4$, $P_2O_5$.

The FCC phase corresponds to the steel substrate. It is naturally registered with ω = 1.15° and it is still visible at ω = 0.70° as it is a well-crystalline phase and the total information depth even exceeds 7 μm at this incidence angle. The same case applies to η-$Al_2O_3$ which was the best fit choice among known transition alumina phases judging by the positions of observable reflections. Concerning the similar decrease of peaks' intensities originating from the steel substrate and corresponding to the η-$Al_2O_3$ phase upon decrease of ω from 1.15° to 0.70° we conclude that the η-phase formed at the steel-$Al_2O_3$ coating interface. This is where the a-few-hundred-nanometre thick layer with columnar grain morphology was identified by TEM (see Fig. 6).

The diffraction pattern collected with the lower ω include also several other considerable peaks besides the ones corresponding to the steel-alumina coating interface (See Fig. 7b). TEM-EDX measurement suggests that the layer is phosphorus-rich, so one plausible explanation for this phenomenon is that, due to the temperature gradient, phosphorus migrated from the steel substrate to the surface of the ceramic coating during exposure to high temperatures. This could result in fluctuations in chemical composition, leading to local variations in thermodynamic properties that, at certain temperatures, created favourable conditions for the crystallization of phosphorus compounds. However, as indicated in Fig. 7 there are other possible phases that can be assigned to existing peaks. Thus, despite the efforts made, a definitive identification of the crystalline layer remains elusive. It should be emphasized that since the thickness of the layer was well below 100 nm, its contribution to the global material mechanical response is expected to be minimal (see Fig. 3a – indentations were made up to 300 nm depth, while the material response is registered from 10x higher volume). Consequently, in light of its apparent lack of impact on the observed mechanical behaviour, we have deemed the precise nature of this crystalline layer as not central to the primary focus and conclusions of this study. While its presence remains an intriguing aspect of our findings, we believe that its omission from detailed discussion does not compromise the overall integrity and significance of the scientific narrative presented in this paper.

### 3.3. *In-situ* high-temperature X-Ray diffraction

The crystal structure evolution of the coating was studied *in-situ* up to 1050 °C using grazing incidence X-ray diffraction. An overview of the collected diffraction patterns (DPs) is presented in Fig. 8. In the Figure, only one diffraction pattern is presented for each temperature step. The patterns shown represent the specimen at the equilibrium state, i.e., any dynamic processes, like recrystallization, that could be observed during the first hours at a particular temperature are not visible in Fig. 8. This also means that, for a given temperature, further soaking didn’t result in any phase transitions. The only exception here is 700 °C, for which two DPs are included in Fig. 8. This aims to show the onset of crystallization.

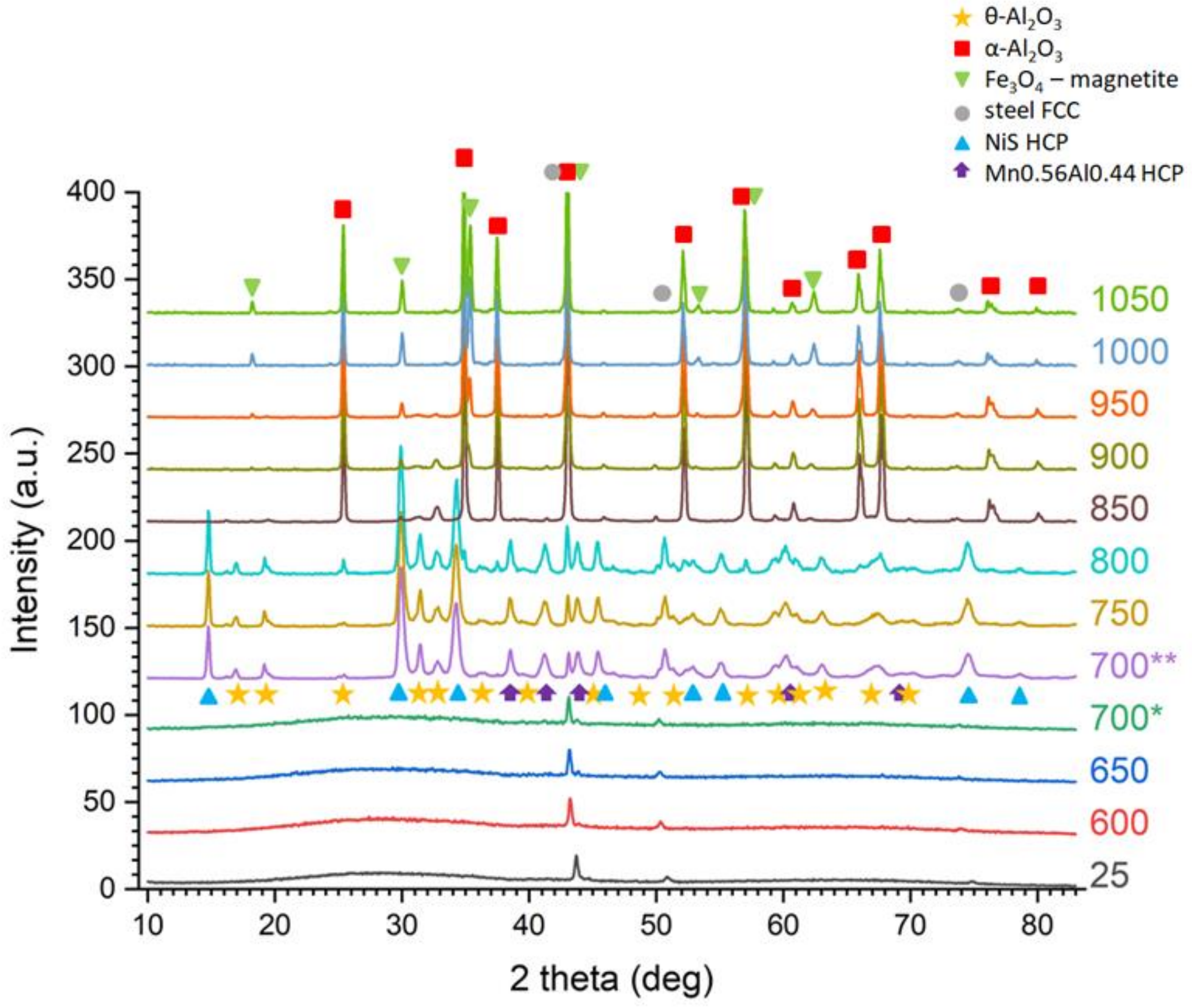

Fig. 8. The overview of the collected diffraction patterns (DPs) representing the specimen's structure at equilibrium or before(*) and after (**) a phase transition at the studied temperatures. Identified crystal phases are marked with symbols, as indicated in the legend in the upper right corner. The numbers on the right represent the temperatures of collecting subsequent DPs.

The detailed diffraction patterns evolution of the coating at 700 °C is shown in Fig. 9. On the map, the onset of the crystallization process was captured. It can be clearly seen that initially, the alumina layer consists exclusively of the amorphous phase. The crystallization from the amorphous structure begins approximately after 15-17 hours of dwelling at 700 °C. The crystallization process continues for 6-8 hours until the crystal structure of the layer becomes stable. After that time, no recrystallization was observed during the next 35 hours.

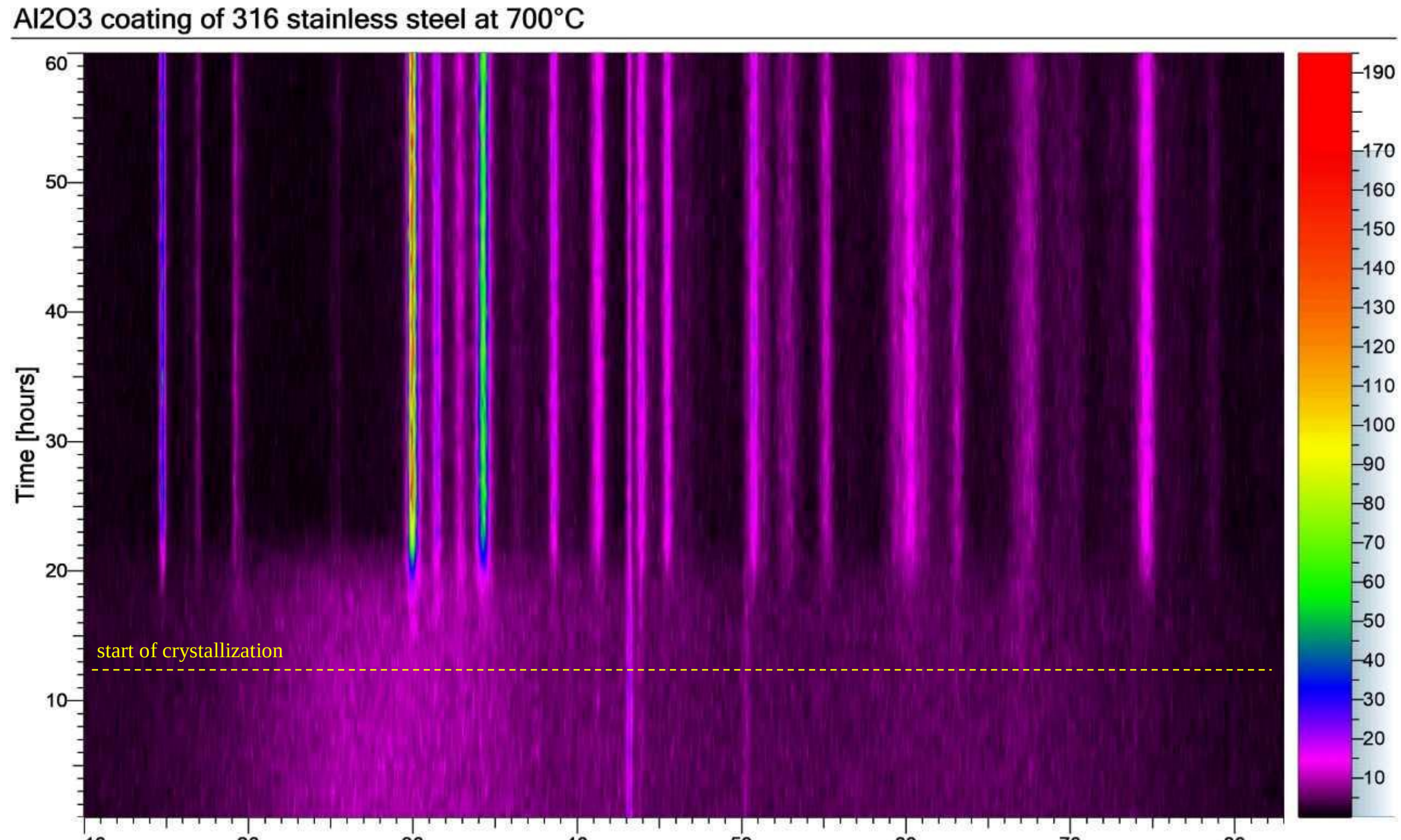

Fig. 9. The overview of the DPs acquired at 700°C; the intensity is marked in color scale from deep purple to orange, as shown on the right. The zero of the time axis is referenced at the beginning of the acquisition of the series of the GID patterns that started 3 h after reaching 700°C. The approximate start of the crystallization is indicated with the yellow dashed line.

| Phase / Temperature | θ-$Al_2O_3$ Monoclinic | α-$Al_2O_3$ Corundum | $Fe_3O_4$ Oxidised substrate | Steel 316 Substrate | NiS HCP | Mn0.65Al0.44 HCP |
|---|---|---|---|---|---|---|
| 25 °C | - | - | - | + | - | - |
| 600 °C | - | - | - | + | - | - |
| 700 °C | + | weak | - | + | + | + |
| 750 °C | + | weak | - | weak | + | + |
| 800 °C | + | + | - | weak | + | + |
| 850 °C | + | + | weak | weak | - | - |
| 900 °C | + | + | weak | weak | - | - |
| 950 °C | weak | + | + | weak | - | - |
| 1000 °C | - | + | + | weak | - | - |
| 1050 °C | - | + | + | weak | - | - |

Fig. 10. The qualitative summary of crystal phases identification at each temperature. Symbols: "-" ≡ the phase is not observed; "+" ≡ the phase is observed; "weak" ≡ selected diffraction peaks of weak intensity corresponding to the phase are present.

The summary of the crystal phases detected in the layer as a function of temperature is presented in Fig. 10. As shown, the alumina layer maintained its amorphous character until 650°C (inclusive). Then, after several hours spent at 700°C, the parasitic reflections originating from the steel substrate became dominated by a couple of new phases. The hitherto amorphous $Al_2O_3$ crystallized in the low-symmetry C2/m space group (no. 12) with the β angle around 104°, usually labelled as the θ phase. Additionally, the corundum polymorph started to crystallize already at 700°C yielding initially very weak peaks that gained intensity as the temperature was increased gradually up to 1050°C. The remaining peaks could be indexed using 2 additional, but separate, structures based on the hexagonally closed packed (HCP) unit cells (space group no. 194, $P\frac{6_3}{m}mc$ and substantially different lattice parameters: $a_1 \approx 3.5$ Å $c_1 \approx 5.4$ Å and $a_2 \approx 2.7$ Å $c_2 \approx 4.4$ Å) that did not represent other $Al_2O_3$ phases. The phase composition identified in the diffraction patterns remained qualitatively stable in temperatures ranging from 700°C to 800°C (inclusive), and only minor effects like thermal expansion and slow crystal growth could be observed. Starting from 850°C, the corundum α-Al2O3 phase dominated the DPs until the specimen reached 1050°C. Corundum was accompanied by the $Fe_3O_4$ magnetite diffraction peaks. The θ-$Al_2O_3$ phase created first at 700°C, could be still recognized at 850°C and 900°C despite its progressive transformation to corundum.

The compositions of the two HCP phases identified in the diffraction patterns at 700-800°C remain largely unknown. These phases might originate from the steel components as the corresponding peaks disappear from the diffraction pattern exactly when $Al_2O_3$ transforms almost quantitatively into corundum and when metal oxides are formed. Metal sulphides, like NiS (e.g. PDF 04-003-4566 [45]), adopt a similar HCP crystal structure, namely the one with the bigger volume of the unit cell ($a_1 \approx 3.5$ Å $c_1 \approx 5.4$ Å) out of the 2 indexed phases. Such metal sulphides are metastable and are known to form around 700°C (1000 K), which is the case in our experiment [59]. The second HCP structure (based on $a_2 \approx 2.7$ Å $c_2 \approx 4.4$ Å) is frequently reported for alloys containing aluminum, like $Mn_{0.56}Al_{0.44}$ (PDF 04-013-9962). Ni, Mn and Sulphur are all components of the 316 steel. Hence, most probably, the discussed HCP phases contain any suitable metallic elements and sulphur from the steel alloy formed at the $Al_2O_3$ layer-steel interface.

### 3.4.High-temperature nanoindentation - Molecular Dynamics (MD) simulations

The molecular dynamics (MD) simulations were performed to gain insight into the atomic-scale behavior of the material under varying temperature conditions. The model outputs are presented in Fig. 11 and Fig. 12.

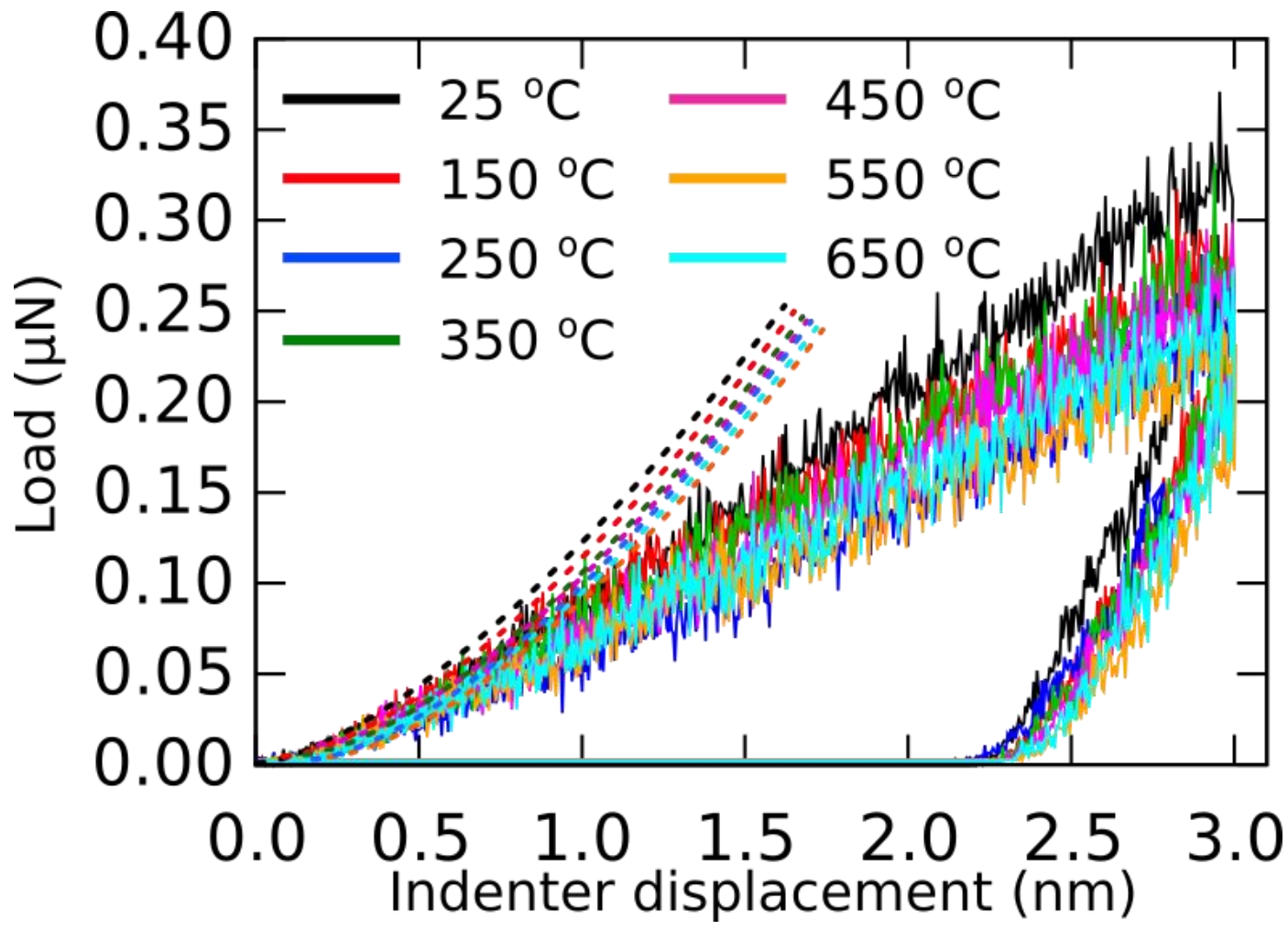

Fig. 11. The MD load vs. depth curves for amorphous alumina under nanoindentation. The Hertz model fitted for the L-D curve representing all measurement temperatures is shown using dashed curves.

Fig. 11. shows the mechanical response of amorphous alumina investigated at temperatures up to 650 °C. A Young's modulus of amorphous alumina was calculated to be about 170, 172, 171, 171, 170, 170, and 170 GPa for 25, 150, 250, 350, 450, 550, and 650 °C, respectively. Simulations performed under varying temperature conditions consistently demonstrate that Young's modulus of alumina remains constant across the temperature range studied. These results are in excellent agreement with experimental data, further supporting the accuracy and reliability of the simulations presented herein. The constant Young's modulus suggests that alumina exhibits a robust mechanical behavior, maintaining its stiffness and elastic properties even under different thermal conditions.

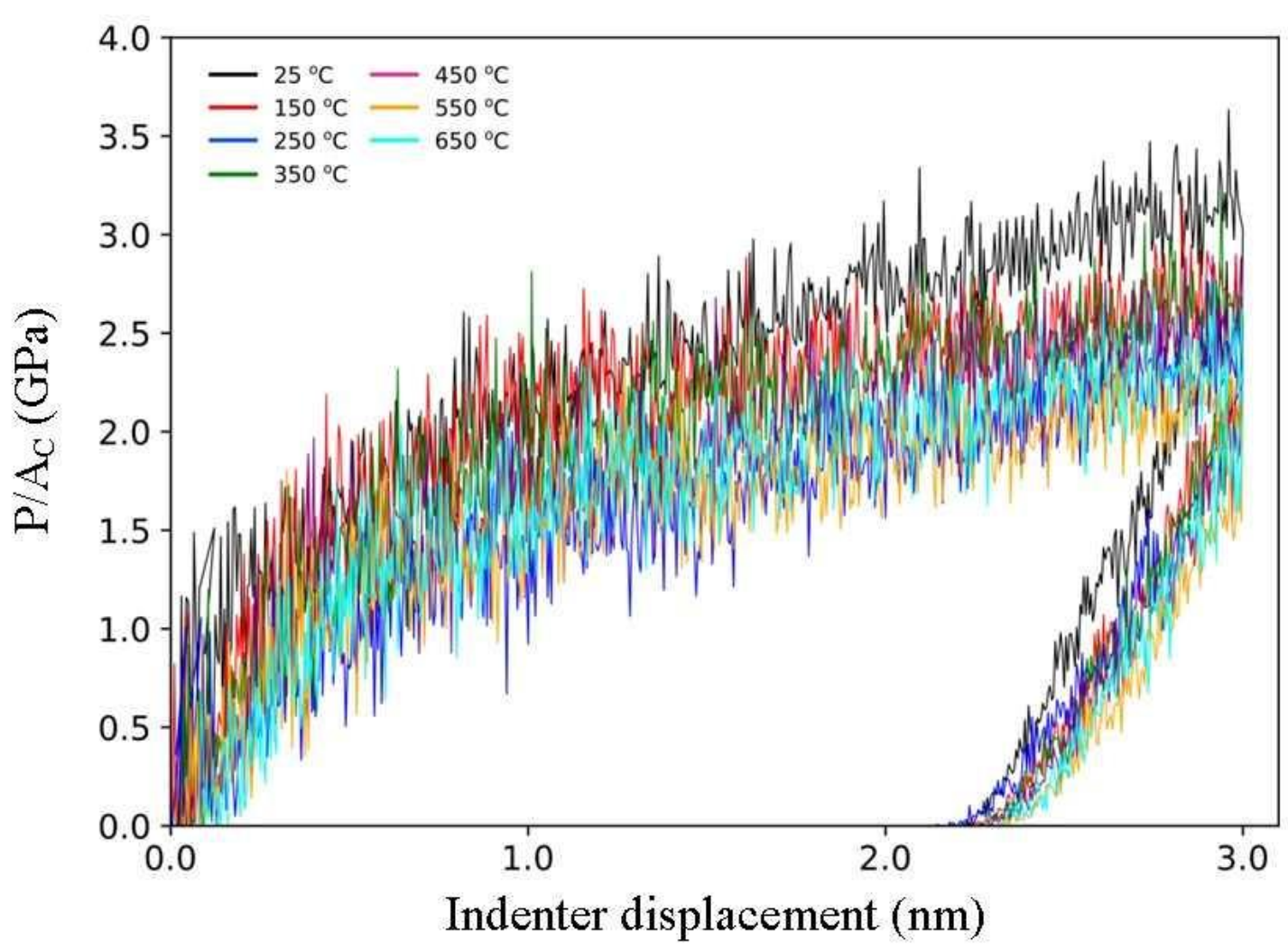

Fig. 12. Contact Pressure as a Function of Indentation Depth

Fig. 12. shows the contact pressure vs. depth curves for the same sample size. All samples were deformed up to δ=3 nm. The hardness of amorphous alumina was calculated to be about 3.17, 2.79, 2.41, 2.63, 2.54, 2.20, and 2.06 GPa for 25, 150, 250, 350, 450, 550, and 650 °C, respectively (please refer to the Table 1 below). MD simulations revealed a decreasing trend in the contact pressure of alumina as the temperature increased, which aligns with prior expectations. The quantitative values obtained from the simulations do not precisely match the experimental measurements and the thermal noise cannot be filtered.

However, the results demonstrate a decreasing trend in contact pressure with increasing temperature, consistent with expectations for amorphous alumina samples. This emphasizes behavioral trends over direct quantitative comparison with experimental data, which is limited by the higher loading rates in simulations.

The goal of the MD simulations was to explore these trends rather than exact experimental replication, enabling valuable insights into the deformation behavior of alumina under varying thermal conditions.

## 4. Discussion

In summary, we have extensively explored the operational properties of the PLD-grown alumina coatings using *in-situ* high-temperature characterization techniques. The mechanical investigation showed that the hardness of the amorphous coating decreases with the temperature. The reported phenomenon is of vital importance, as from the mechanical point of view, the alumina coating-steel substrate system proves to be well-matched at high temperatures. To complete the picture, the mechanical results were interpreted in the light of the structural analysis. Both post-mortem characterization of the HT indented sample and *in-situ* HT study on the as-deposited coating lead to the conclusion that the softening phenomenon observed in this study is related to the material's behavior in its amorphous state. These results widen our knowledge of the high-temperature behavior of aluminum oxide, as the previous research was limited to the investigation of crystalline hcp polymorphs of $Al_2O_3$. The data published by M. Rebelo de Figueiredo et al. [35] show that a reduction of the hardness and reduced modulus with temperature is observed for α-and κ-$Al_2O_3$ coatings synthesized by chemical vapor deposition (CVD). These findings match well with the data available in the literature for the bulk single-crystal and polycrystalline sapphire [52–55,60,61]. Unexpectedly, our study revealed that Young's modulus of the amorphous alumina is constant with temperature. Typically, as temperature increases, the mechanical characteristics tend to decline due to lattice vibrations causing a decrease in bond strength. However, Young's modulus, a measure of bond strength, appears to remain constant within the temperature range being studied in the experiment. This suggests that a different mechanism is responsible for the reduction in hardness. To some extent, the registered effect of the temperature may be explained in terms of stress relaxation. It has been shown in the literature that even amorphous coatings can exhibit high levels of residual stress [62–64], which can affect their performance. It is unavoidable for residual stresses to be imparted onto the coating during deposition, while deposition temperature can significantly impact their magnitude in amorphous coatings. As the investigated coating was grown at room temperature, the level of compression stresses in the as-deposited state is expected to be relatively high. Nevertheless, since the MD simulation (that assumes that the material is unstressed) indicates a consistent trend, the softening cannot be explained only in terms of stress relaxation. The observed decrease in hardness with increasing temperature poses an intriguing phenomenon requiring further investigation to fully comprehend its underlying physical mechanisms. Despite the comprehensive analysis conducted in this study, the exact reason for this temperature-dependent decrease in hardness remains elusive. Based on the literature review, we propose that the observed softening phenomenon can be explained in terms of thermal activation that allows sufficient bond-switching and, thus, plasticity, which is supported by the study by Erkka J. Frankberg et al. [36]. Recently, the same team proved that plasticity does not occur only in the nanoscale but transcends to the microscale (proved by micro-pilar compression which deformed without fracture at up to 50% strain via a combined mechanism of viscous creep and shear band slip propagation) [36]. Conducted atomistic simulations confirm that the bond switch-driven nucleation of localized plastic strain events remains highly active in amorphous alumina. This effect has been observed at different strain rates. Frankberg et al. [65] suggested that the nucleation occurs at a fast enough rate to play an important role in preventing the catastrophic propagation of the dominant shear band slip, which is a known fracture mechanism in brittle metallic glasses [66,67]. Our data supports this study and suggest that this mechanism is accelerated at high temperature, further proving that amorphous materials can have a ductile response. Regardless, the main conclusion drawn from the high-temperature mechanical investigation suggests that the coating may exhibit enhanced performance under operating conditions, indicating potential improvements compared to room-temperature conditions.

The experimental mechanical research has been complemented and supported by the simulations. Tab. 1 below compares nanoindentation results, specifically hardness, and Young's modulus, obtained from simulations by Eq. 4 and experimental measurements. The data enables a direct assessment of the agreement between simulated and experimental values, providing insights into the accuracy and reliability of the data obtained by two independent routes. The absolute values of Young's modulus were the same (no more than 10% difference), indicating that the simulations accurately predicted the material's stiffness at different temperatures. Moreover, Young's modulus with temperature trend exhibited consistency between the experimental and simulated data. These findings highlight the reliability of the simulation model in accurately predicting the temperature-dependent behavior of the material's stiffness and indicate that the experimental methodology was correctly selected. On the other hand, when examining hardness, there were disparities between the absolute values obtained from simulations and experimental measurements. This observation implies that the model is unsuitable for accurately predicting the absolute hardness values at different temperatures. However, despite the discrepancy in absolute values, the trend of hardness with temperature remained consistent between simulations and experiments. It is widely acknowledged and frequently observed that simulations do not yield absolute values that match experimental data in various scientific studies [68,69].

In conclusion, this comprehensive analysis of nanoindentation results highlights the agreement between experimental and simulated values for Young's modulus in terms of both absolute values and temperature-dependent trends. Although there were variations in the absolute hardness values between simulations and experiments, the consistent trend observed for hardness provides confidence in the simulation's ability to capture the relative changes with temperature. These findings emphasize the importance of considering experimental and simulated data in comprehensive studies of material behavior.

Tab. 1. Comparison of Nanoindentation Results: Simulations vs. Experimental Data

| Temperature (°C) | Hardness (GPa) | | Young's modulus (GPa) | |
|---|---|---|---|---|
| | Experiment | MD simulation | Experiment | MD simulation |
| 25 | 9.6±0.4 | 3.2 | 175±5 | 157 |
| 150 | 7.4±0.2 | 2.8 | 173±4 | 154 |
| 250 | 7.3±0.2 | 2.4 | 180±3 | 158 |
| 350 | 6.5±0.2 | 2.6 | 172±7 | 158 |
| 450 | 6.5±0.3 | 2.5 | 203±27 | 157 |
| 550 | 5.2±0.2 | 2.2 | 169±24 | 156 |
| 650 | 4.8 | 2.1 | 172.0 | 157 |

In addition to experimental and computational mechanical research, high-temperature structural tests were conducted further to investigate the material's response under elevated temperature conditions. The findings of the structural study show that the amorphous-crystal transition temperature for the PLD-grown amorphous alumina lies between 650 and 700 °C, which is above the foreseen operating conditions in LFR technology. Nevertheless, the radiation-induced crystallization of the initially amorphous alumina results from the synergetic effect of high temperature and radiation damage [70]. Hence, the higher the irradiation level, the lower the temperature threshold for the onset of crystallization. The unique properties of PLD-grown alumina coating, which make it an extremely attractive material for the nuclear industry, largely follow from its amorphous structure. The characteristic brittleness does not limit its usefulness as in the case of bulk sapphire [36]. Even a minimal risk of breaking off the coating and eventually exposing the unprotected substrate or blocking the coolant channels is unacceptable in nuclear reactors. Previous results demonstrate that PLD-grown $Al_2O_3$ coatings can retain structural integrity and adhesion upon irradiation-induced crystallization [22], so there is no clear evidence that coating crystallization could inevitably lead to the system's brittle failure. In addition to that, our study shows that amorphous alumina oxide increases its plasticity with temperature.

Nevertheless, the control of phase transformations seems to be critical to guarantee the maximum stability of the system's performance. This is due to the fact that occurring continuous evolution of a material properties makes the prediction of its behavior at a given moment very challenging. The issue becomes particularly acute when the changes may result in material degradation that could pose a risk to the system's reliability and safety. An effective solution that could notably tackle the issue of performance instability consists of doping, i.e., injecting the predefined element into the material structure. Much work has been carried out on the potential stabilization effect [71–74]. It has been shown that crystallization onset can be successfully suppressed in this way. The strategy is considered a prospective direction in the area – the design and optimization of amorphous element-stabilized alumina coating fabrication will be a challenge for the following years.

## 5. Summary

This paper reports for the first time an in situ high-temperature study of thin amorphous alumina coatings grown by PLD for nuclear applications. Implementing HT GIXRD allowed us to monitor the temporal evolution of phase transitions in the material at temperatures exceeding 650 °C. It has been shown that a thermally activated amorphous-crystalline transition occurs at 700 °C. Furthermore, we observed the intermediate alumina θ-$Al_2O_3$ phase up to 950 °C, with the exclusively thermodynamically stable α-$Al_2O_3$ becoming the dominant phase at higher temperatures. The paper highlights the importance of phase transformation control in amorphous alumina, as it is critical to the performance of the system and a crucial point for understanding the damage behavior of such coatings. HT Nanoindentation measurements enabled analysis of the stability of the system's mechanical performance up to 650 °C. It has been demonstrated that as the temperature increases, the amorphous alumina experiences a constant gradual decrease in hardness and an increase in plasticity. The experimental findings correlate well with the output of MD models. This constancy suggests that the primary bonds in the material retain their strength even at elevated temperatures. In contrast, the observed decrease in hardness with temperature reflects an increase in plasticity, which is governed by bond-switching mechanisms rather than bond strength. Recent studies (Erkka J. Frankberg et al. [36,65]) have shown that plastic deformation may occur in amorphous materials like alumina through bond switching - where atoms retain their local coordination environment but shift their bonding neighbors. This process allows deformation without significant density changes, accommodating both shear and tensile stresses. The breaking and reforming of bonds during this mechanism accelerate at higher temperatures, explaining the increased plasticity that is observed in our study. Since Young's modulus is tied to the stiffness of individual bonds, it remains constant because the overall bond strength is unaffected by the switching mechanism.

This consistency supports the reliability of our findings. The phenomenon that has been reported is crucial since it indicates that the alumina coating-steel substrate system is mechanically well-suited for use at high temperatures. Our study shows that high-temperature tests can help understand material behavior in high-temperature applications and design coating systems for enhanced performance. We believe that our research will serve as a base for future studies on enhanced high-temperature performance element-stabilized alumina.

Acknowledgments

The research leading to these results was carried out in the frame of the EERA Joint Programme on Nuclear Materials and is partly funded by the European Commission Horizon 2020 Framework Programme under grant agreement No. 755269 (GEMMA project) and grant agreement No. 101061241 (INNUMAT). We acknowledge support from the European Union Horizon 2020 research and innovation program under grant agreement no. 857470 and from the European Regional Development Fund via the Foundation for Polish Science International Research Agenda PLUS program grant No. MAB PLUS/2018/8. Also, financial support from the National Centre for Research and Development through a research grant,“"Studies of the role of interfaces in multi-layered, coated and composite structures”" PL-RPA2/01/INLAS/2019, is gratefully acknowledged. Lastly, financial support from the Ministry of Science and Higher Education was granted under agreement no. 3906/H2020-EURATOM/2018/2 is gratefully acknowledged.